\documentclass[prl,showpacs,showkeys,twocolumn,a4paper,amsmath,floatfix]{revtex4-1}
\pdfoutput=1

\usepackage{graphicx}
\usepackage{dcolumn}
\usepackage{bm}
\usepackage{color,soul}
\usepackage{amsmath}

\begin{document} 

\title{Methods to calibrate and scale axial distances in confocal microscopy as a function of refractive index}

\author{T. H. Besseling}
\email{t.h.besseling@uu.nl}
\affiliation{Soft Condensed Matter, Debye Institute for
NanoMaterials Science, Utrecht University, Princetonplein 1,
NL-3584 CC Utrecht, the Netherlands}

\author{J. Jose}
\affiliation{Soft Condensed Matter, Debye Institute for
NanoMaterials Science, Utrecht University, Princetonplein 1,
NL-3584 CC Utrecht, the Netherlands}

\author{A. van Blaaderen}
\email{a.vanblaaderen@uu.nl}\homepage{http://www.colloid.nl}
\affiliation{Soft Condensed Matter, Debye Institute for
NanoMaterials Science, Utrecht University, Princetonplein 1,
NL-3584 CC Utrecht, the Netherlands}

\date{\today}

\begin{abstract}
Accurate distance measurement in 3D confocal microscopy is important for quantitative analysis, volume visualization and image restoration. However, axial distances can be distorted by both
the point spread function and by a refractive-index mismatch between the sample and immersion
liquid, which are difficult to separate. Additionally, accurate calibration of the axial distances in confocal microscopy remains cumbersome, although several high-end methods exist. In this paper we present two methods to calibrate axial distances in 3D confocal microscopy that are both accurate and easily implemented. With these methods, we measured   axial scaling factors as a function of refractive-index mismatch for high-aperture confocal microscopy imaging. We found that our scaling factors are almost completely linearly dependent on refractive index and that they were in good agreement with theoretical predictions that take the full vectorial properties of light into account. There was however a strong deviation with the theoretical predictions using (high-angle) geometrical optics, which predict much lower scaling factors. As an illustration, we measured the point-spread-function of a point-scanning confocal microscope and showed that an index-matched, micron-sized spherical object is still significantly elongated due to this PSF, which confirms that single micron-sized spheres are not well suited to determine accurate axial calibration nor axial scaling.
\end{abstract} 
 
\keywords{confocal microscopy, axial calibration, axial scaling, refractive index mismatch} 
 
\maketitle
\section{Introduction}
Confocal microscopy is a powerful tool for 3D in-situ measurements of both structure and dynamics for a wide range of scientific disciplines, such as cell-biology, pharmaceutics and materials science \citep{Wilson1990,Pawley2006a,White2005,Prasad2007}. However, care has to be taken with 3D measurements because not all three dimensions are effected in the same way by both optics and data acquisition software. The inevitable difference in lateral and axial resolution  affects the apparent shape of any (sub)micron-sized feature in a 3D measurement \citep{Jenkins2008}. Furthermore, there is often a refractive index (RI) mismatch between immersion fluid and sample. Not only does the RI-mismatch deteriorate the point spread function (PSF) with increasing focus depth, and therefore the resolution, it also introduces a decrease in intensity and a shift of the objective focus \citep{Visser1992,Hell1993,Visser1994,Pawley2006,Sheppard1994,Sheppard1997,Wiersma1997,deGrauw1999,Diaspro2002,Neuman2005,Shaevitz2007}.   When the refractive index of the sample is smaller than the immersion liquid used for imaging, axial distances appear more elongated due to the refractive effects on the focus position. A clear distinction can be made between studies that analyse these focal shifts with geometrical optics and studies that take the  vectorial properties of light into account. On the basis of geometrical optics, axial elongation up to a factor of three times the actual distance has been predicted for high-aperture oil-immersion imaging in aqueous samples \citep{Visser1992,Visser1994}. It seems likely however that in the mechanism of the axial shift, paraxial rays dominate over the high-angle rays that are used in the geometrical optics approach \citep{Sheppard1994}. Studies that take the vectorial properties of light into account therefore predict significantly smaller axial elongations \citep{Hell1993,Jacobsen1995,Sheppard1997}.
\\ \indent There are however still significant differences between the precise values of the axial scaling factors for different vector-based theories \citep{Hell1993,Jacobsen1995,Sheppard1997,Wiersma1997} and the amount of experimental studies remains limited \citep{Hell1993,White1996,Neuman2005}. Also, in most experimental studies on axial distance scaling, little attention is devoted to the axial-distance \textit{calibration}, which is indispensable for precise measurements. Calibration of the lateral distances is both straightforward and accurate, e.g.\ by using a calibration grid. However, accurate calibration of the axial distances in confocal microscopy remains cumbersome, although several high-precision methods exist, e.g.\ using a focus function \citep{Boddeke1997} or via aggregates of colloidal spheres with a narrow size distribution \citep{Bornfleth1998}. \\ \indent In this paper we demonstrate two methods to calibrate axial distances in confocal microscopy that are both accurate and practical to employ. In the first method we use light interference to accurately measure the height of an empty calibration cell. We filled the cell with four different solvents mixed with fluorescent dye, which enabled the determination of the axial scaling factors as a function of refractive index for high-aperture 3D confocal-microscopy imaging with an oil-immersion objective. We also demonstrate a second method to accurately calibrate the confocal microscope, which is with large ($\sim$~50 $\mu$m)  spherical particles that only have a thin fluorescent shell (compared to their size). Finally, we show as an illustration of our $z$-calibration a measurement of the PSF of a confocal microscope and we demonstrate that a single, spherical object of $\sim$~1 $\mu$m is still significantly elongated due to this PSF, even in the absence of a RI-mismatch.

\section{Methods}
\subsection{Calibration cell construction and FTIR measurement}
To calibrate the axial distances in a point-scanning confocal microscope, we built a custom sample cell with standard glass coverslips (Menzel Gl\"{a}zer). The glass coverslips had a refractive index ($n_D^{23} = 1.523$) close to the refractive index of the oil-immersion liquid (Type F, Leica, $n_D^{23}=1.515$) used for imaging. We avoided using glass capillaries (Vitrocom), often used in confocal studies on colloidal systems, since they provide lower quality   imaging which is partially due to their manufacturing process and also due to the refractive index ($n_D^{23}$ = 1.47). We used a standard No.~1.0 coverslide, which has a thickness between 130 - 160 $\mu$m, as specified by the manufacturer (Menzel Gl\"{a}zer). Although standard confocal microscopy objectives are optimized for a coverslip thickness of 170 $\mu$m \citep{Pawley2006a} and therefore a No.~1.5 coverslip (thickness 160 - 190 $\mu$m) would have been more accurate, we could not however completely image our cell (with a height $\sim$~80 $\mu$m), due to the limited working distance of the high numerical aperture objectives that we used. As spacers, we used No.~00 coverslips (thickness 55 - 80 $\mu$m) and the individual components of the cell were permanently fixed onto a standard microscopy slide (Menzel Gl\"{a}zer) with UV glue (Norland 68 Optical Adhesive). The resulting height of the cell $H$ was measured with a Fourier Transform Infrared (FTIR) spectrometer (Vertex 70, Bruker). To avoid additional interference effects from the top coverslip itself, a drop of immersion oil was carefully placed on top of the cell before the measurement. The thickness and irregularities of the much thicker microscopy slide ($\sim$ 1 mm) made it not necessary to correct for its interference effects.

\subsection{50 $\mu$m PMMA spheres}
We used large poly(methyl methacrylate) (PMMA) spheres as a second method for calibration. The spheres had an average diameter $\sigma$ = 50 $\mu$m and polydispersity $>$ 10\% (Altuglas, BS150N). To fluorescently dye the particles, we first prepared (rhodamine isothiocyanate)-aminostyrene (RAS) dye following the method described by Bosma  \textit{et al.}\ \citep{Bosma2002}. Then, we saturated a quantity of acetone (99\%, Merck) with RAS and subsequently centrifuged the saturated acetone at high speed to sediment undissolved dye. The acetone was then added to dodecane (99\%, Sigma-Aldrich) to give a 10 wt\% solution of acetone. In this mixture, 50 wt\% undyed PMMA particles and 0.35 wt\% azo-bis-isobutyronitrile (98\%, Acros) were suspended in a glass vial. The reaction mixture was heated up to 83$^\circ$C and left to react for approximately 1 day. During this reaction, RAS molecules become chemically bonded with unreacted PMMA-ends at the surface of the particle. The vial was left open, so acetone could evaporate. The dyed particles were washed with hexane and dried under vacuum. Afterwards, the particles were suspended in a 24 wt\% mixture of cis-decahydronaphthalene (cis-decalin, 99\%, Sigma-Aldrich) in cyclohexylbromide (CHB, 98\%, Sigma-Aldrich). The refractive index of this mixture was  $n_D^{21}$ = 1.490, as measured with a refractometer (Atago 3T). This solvent mixture closely matched the refractive index of the particles, based on the fact that the refractive index is close to that of the bulk material ($n_D^{20}$ = 1.491 \citep{Kasarova2007}) and that the particles hardly scattered when viewed under bright-field illumination.

\subsection{Confocal microscopy measurements}
The confocal microscopy measurements were all performed with a Leica SP2 or Leica SP8. All distance measurements were performed on 3D image stacks obtained in \textit{xyz}-scanmode. Although a (single) vertical scan obtained in \textit{xzy}-mode is a fast method to view vertical slices through the sample, the obtained distances are in general not accurate and were avoided for any quantitative measurement. Imaging of the empty calibration cell was performed with a 20x/0.7 air-objective (Leica), all other measurements were performed with a 100x/1.4 oil-immersion confocal objective (Leica). The largest measurement error is introduced by the top coverslip being under a small angle with respect to the microscopy glass slide (see Fig.~\ref{fig:cell}a), despite careful application of the UV glue. Because we cannot place the sample in exactly the same position after its first measurement, we measured the height gradient in the \textit{x}- and \textit{y}-direction and found that the largest slope was 1.9 $\mu$m/mm. Assuming that it is possible to place the sample in its original position within 0.3 mm accuracy, a rough estimate of the error on the confocal height measurements is $\sim$ 0.6 $\mu$m. We therefore chose our pixel-size in the axial direction to roughly half of this value. For the axial-scaling measurements, we used solvents of increasing RI: immersion oil (Type F, Leica, $n_D^{20} = 1.516$), cyclohexylchloride (CHC, $>$98\%, Merck, $n_D^{20}$ = 1.463), dodecane ($>$99\%, Sigma-Aldrich, $n_D^{20} = 1.421$) and de-ionized water (Millipore system, $n_D^{20} = 1.333$). The first three (apolar) solvents were saturated with pyrromethene-567 dye (excitation maximum $\lambda_{max} = 518$ nm, Excition) whereas the water was saturated with fluorescein isothiocyanate (FITC, isomer I, 90\%, Sigma-Aldrich). Undissolved dye was removed by centrifugation. Also, a small amount of sterically stabilized PMMA tracer particles \citep{Bosma2002} (diameter $\sigma$ = 2.07 $\mu$m, polydispersity 3\%), that often stick to untreated glass, was added to the apolar solvents to accurately determine the top and bottom of the cell. Because the volume fraction of the PMMA tracer particles is $\ll$ 1 \%, their contribution to the effective refractive index of the sample can be neglected. Solvents were removed from the sample cell with nitrogen flow and the cell was flushed three times with the new solvent before the sample was carefully placed on the marked area under the confocal microscope to record a new image-stack. The image-stacks of the calibration cell were all recorded on a Leica SP2 with a 488 nm laser and a scan speed of 1000 Hz. The voxel-size of the image stacks was 293 x 293 x 311 nm$^3$. The typical total volume of the images stacks was 38 x 38 x 115 $\mu$m$^3$. Images of the large PMMA spheres ($\sigma$ = 50 $\mu$m) were recorded on a Leica SP8 with a 543 nm laser line, voxel-size 51 x 51 x 168 nm$^3$ and total volume 52.8 x 52.8 x 54.1 $\mu$m$^3$.

\subsection{PSF measurement \& deconvolution}
We measured the point spread function (PSF) of a confocal microscope which was subsequently used to deconvolve 3D image-stacks of spherical particles (diameter $\sigma$ = 200 nm and 1040 nm) positioned close to the coverslip. For the deconvolution of the image-stack of the large PMMA sphere ($\sigma$ = 50 $\mu$m), we used a depth dependent theoretical PSF that takes into account the (small) RI-mismatch between sample and immersion fluid. All image restorations were performed using commercially available software (Huygens Professional 4.4, Scientific Volume Imaging) using the classic maximum likelihood estimation restoration method \citep{VanderVoort1995}. 
To measure the PSF, we used fluorescent polystyrene spheres with diameter $\sigma = 200$ nm, polydispersity 5\% and excitation maximum $\lambda$ = 441 nm as bought (YG Fluoresbrite Microparticles, Polysciences). The polystyrene particles (bulk material $n_D^{20}$ = 1.592 \citep{Kasarova2007}) were dried on a cover glass (Menzel Gl\"{a}zer, No.~1.5) and subsequently a drop of immersion oil (Type F, Leica, $n_D^{20}$ = 1.516) was placed on the glass slide to (nearly) index-match the particles. The sample was then placed on a microscopy slide with glass spacers and sealed with UV glue (Norland Optical Adhesive). Images of the beads were recorded with an inverted confocal microscope (Leica SP8) with a 100x/1.4 oil immersion objective (Leica) in combination with a Hybrid detector. To gain enough statistics, confocal image-stacks of 8 different spheres were recorded with (sub)Nyquist sampling rate (18.2 x 18.2 x 83.9 nm$^3$). Because these particles are only approximate point-sources, the PSF was obtained by iterative deconvolution with a 200 nm bead object \citep{VanderVoort1995}. 
Additionally, we imaged poly(methyl methacrylate) (PMMA) spheres with diameter $\sigma$ = 1040 nm and a polydispersity $\delta$ = 3\%, as determined with static light scattering (SLS). The particles were sterically stabilized with poly(12-hydroxystearic acid) (PHS) grafted onto the PMMA backbone which was chemically attached to the core of the particles and covalently labelled with fluorescent 4-methylaminoethylmethacrylate-7-nitrobenzo-2-oxa-1,3-diazol (NBD-MAEM) dye for imaging \citep{Bosma2002}. With the measured PSF, we deconvolved image-stacks of both the fluorescent polystyrene spheres ($\sigma = 200$ nm) and of the larger PMMA spheres ($\sigma$ = 1040 nm) that were dried on a glass coverslip (Menzel Glazer, No. 1.5) and subsequently immersed in immersion oil (Type F, Leica). The particles were imaged within one hour of sample preparation. We acquired images stacks with voxel-size 5.4 x 5.4 x 41.96 nm$^3$ and 18.75 x 18.75 x 83.9 nm$^3$ respectively, using a 100x/1.4 oil objective and a 488 nm laser-line selected from a white light laser.

\section{Results}
\subsection{Calibration cell \& distance measurements}
\begin{figure}[ht!]
\includegraphics[width=0.40\textwidth]{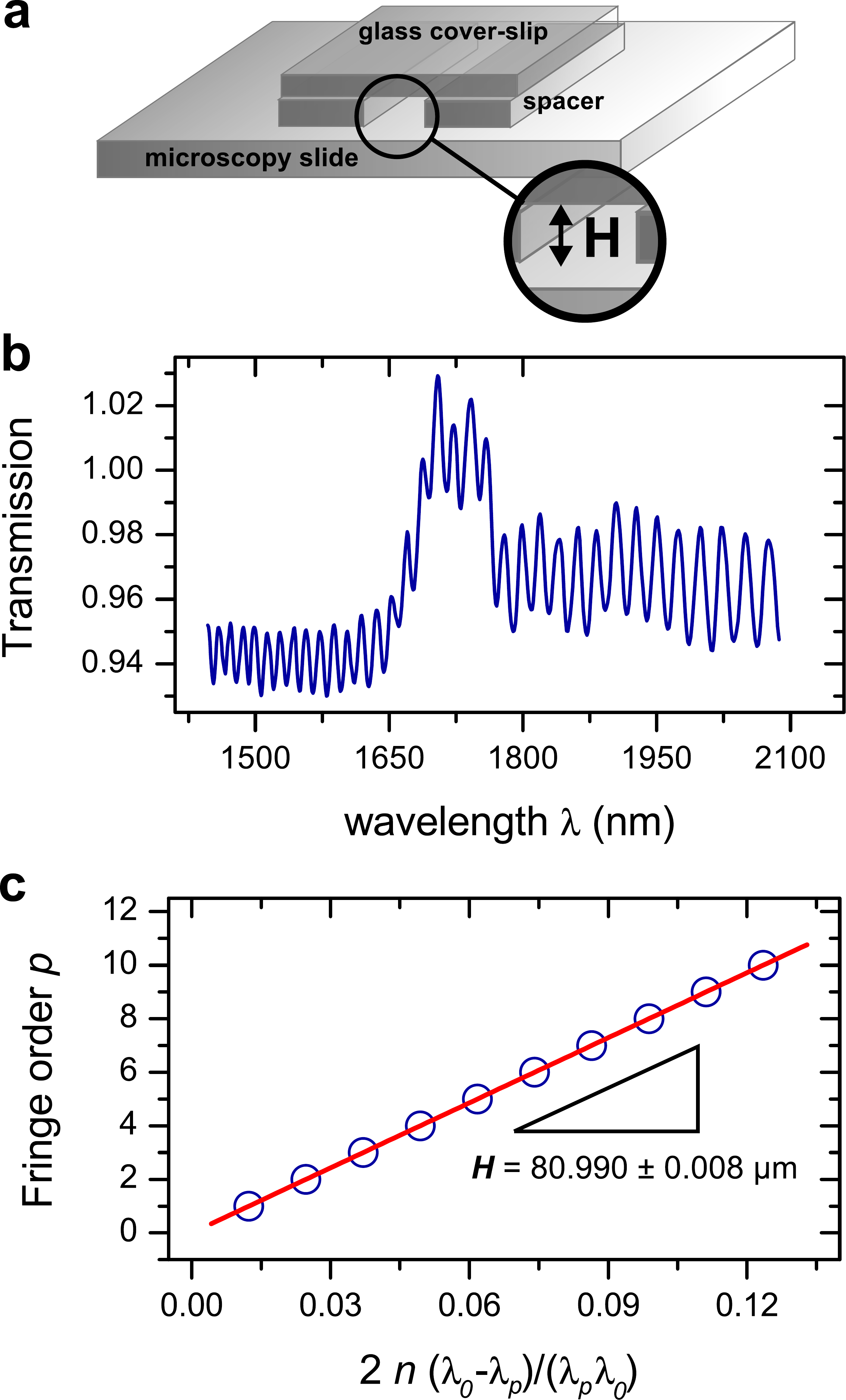}
\caption{Construction and measurement of a calibration cell. 
(a) A sample cell with height $H$ was built with glass coverslips and a standard microscopy slide, glued together with UV-glue. 
(b) When the (empty) cell was placed in a Fourier Transform Infrared (FTIR) spectrometer, Fabry Perrot (FP) fringes were visible in the transmission spectrum. 
(c) The height of the cavity ($H = 80.990 \pm 0.008$ $\mu$m) was determined from the spacing between the FP fringes \citep{Jiang1999a}. The error-bars on individual points are smaller than the symbol size.\label{fig:cell}
}
\end{figure}

The sample cell used for calibration is shown in Fig.~\ref{fig:cell}a. When placed in a spectrometer, light reflecting from the front and back of the inside of the sample cell resulted in oscillations in the transmission spectrum, known as Fabry Perrot (FP) fringes, and shown in Fig.~\ref{fig:cell}b.
We determined the height of the cell from the spacing between the maxima of the FP fringes with the formula \citep{Jiang1999a}
\begin{equation}
	H=p \frac{\lambda_p \lambda_0}{2 n (\lambda_0-\lambda_p)},
\end{equation}
with $\lambda_0$ the longest wavelength, $p$ the fringe order of subsequent maxima at wavelength $\lambda_p$ and $n$ the refractive index of the medium (air). In Fig.~\ref{fig:cell}c the fringe order $p$ is plotted as a function of $2 n (\lambda_0-\lambda_p) / \lambda_p \lambda_0$. The slope of the linear fit directly gives the height of the cell $H = 80.990 \pm 0.008$ $\mu$m.

\begin{figure*}[ht]
\includegraphics[width=0.95\textwidth]{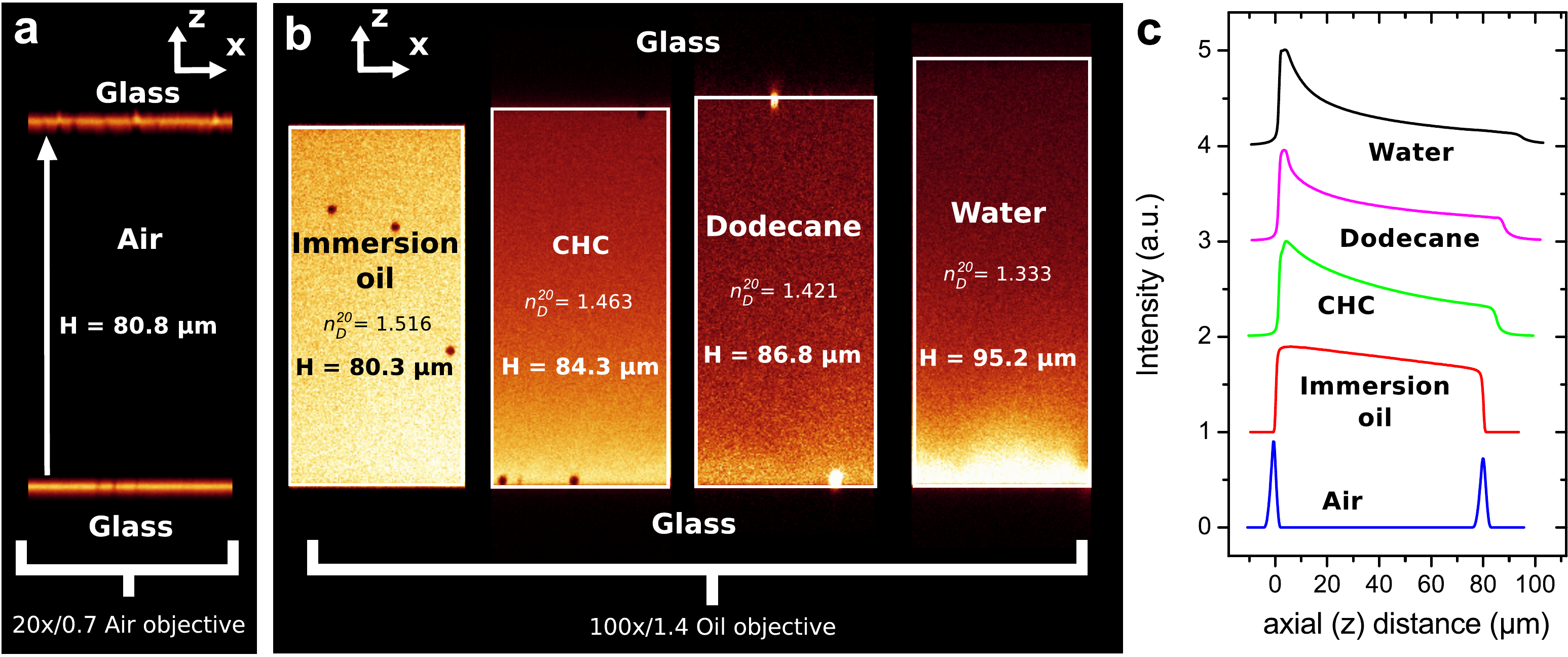}
\caption{Axial distances measured with confocal microscopy. (a) The empty calibration cell with  $H = 80.990 \pm 0.008$ $\mu$m was measured in confocal reflection mode (Leica SP2) with a 20x/0.7 air-objective (Leica), which resulted in $H=80.8 \pm 0.3$ $\mu$m. (b) The cell filled with immersion oil, pyrromethene dye and poly(methyl methacrylate) (PMMA) tracer particles (left). The sample was imaged with an 100x/1.4 oil objective (Leica) and a similar height was measured ($H = 80.3$ $\mu$m). However, when the cell was re-filled with solvents that had a refractive index-mismatch with the oil-objective, deviating axial-distances were found, as indicated in the figure. (c) Intensity profiles along the axial ($z$) direction show the increase in (apparent) axial distance as well as decrease of intensity deeper in the sample. The profiles where normalized and shifted for better visualization.
\label{fig:cell_full}}
\end{figure*}

\begin{figure}[h!]
\includegraphics[width=0.5\textwidth]{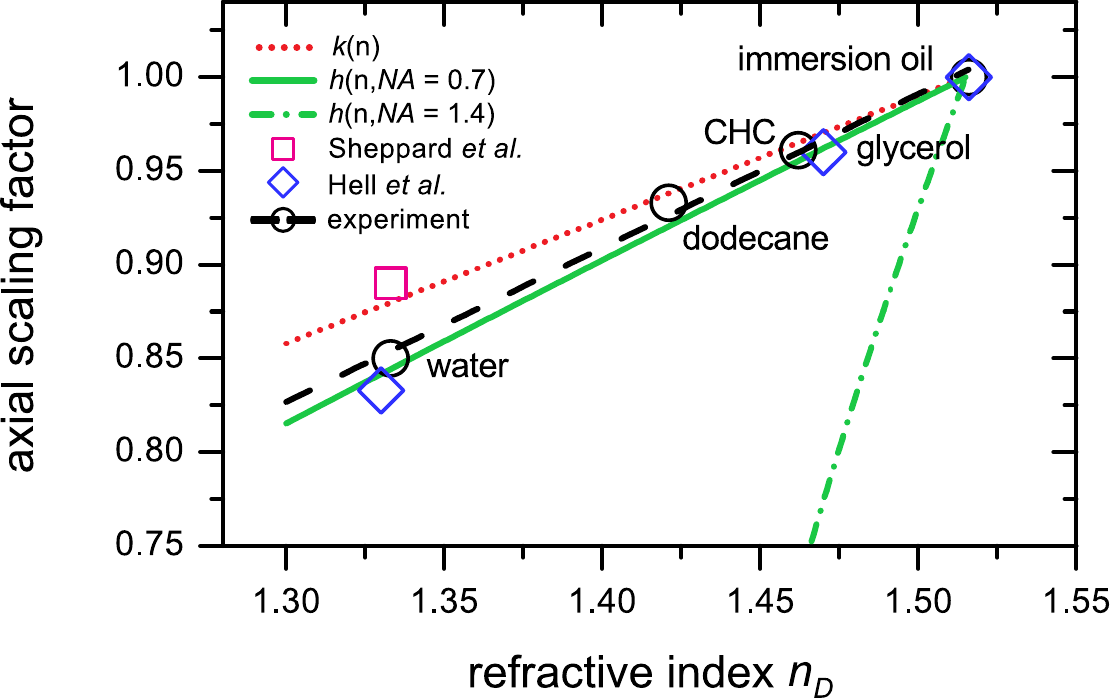}
\caption{Axial scaling factors as a function of the sample refractive index $n_D$. Our measurements are indicated with black open circles, which were fitted with the dashed (black) line. They are in good agreement with the results of Hell \textit{et al.}\ \citep{Hell1993}. They are however in strong contrast with the high-angle geometrical prediction of equation  (\ref{eq:h}) and deviate to a lesser extend with the paraxial limit of equation (\ref{eq:k}).
\label{fig:scaling}}
\end{figure}

In Fig.~\ref{fig:cell_full}a we show a confocal micrograph of the empty calibration cell, imaged in reflection mode with a 20x/0.7 air objective and 488 nm laser. The image clearly shows the reflections at the glass-air interfaces, which we assumed to be positioned at the highest pixel-intensity. We measured the height at the same position as was done with the spectrometer (for four different times), which resulted in a mean value of H = 80.8 $\pm$ 0.3 $\mu$m. This value is in good agreement with the spectrometer measurement ($H = 80.990 \pm 0.008$ $\mu$m) and thus confirms proper calibration of the microscope in the axial direction. \\ \indent Fig.~\ref{fig:cell_full}b shows the same cell, this time filled with solvents of decreasing refractive index, as indicated in the figure. The tracer particles were used to measure the height of the sample. When the cell was filled with immersion oil (Fig.~\ref{fig:cell_full}b, left) a single value of H = 80.30 $\mu$m was obtained. After removal of the oil, the empty cell was measured again with an air objective which resulted in a value of H = 80.92 $\mu$m. From these measurements we can conclude that the confocal was accurately calibrated and that filling the cell with solvent did not alter the height significantly. \\ \indent We also measured the effect of refractive index (RI) on the axial distances, indicated by the intensity profiles shown in Fig.~\ref{fig:cell_full}c. Not only does the (apparent) axial distance change as a function of RI, also the intensity becomes non-linearly  dependent on the axial distance, which is described in detail elsewhere \citep{Hell1993}. We compared the data obtained from Fig.~\ref{fig:cell_full}c with a theoretical model for the scaling factor of axial distances $h(n,NA)$, based on geometrical optics, given by \citep{Visser1992,Visser1994}
\begin{equation}
	h(n,\textit{NA}) = \sqrt{\frac{n^2-\textit{NA}^2}{n_{oil}^2-\textit{NA}^2}}, 
	\label{eq:h}
\end{equation}
with $n$ the refractive index of the suspension, $n_{oil}=1.516$ the refractive index of the oil immersion liquid and NA the numerical aperture of the objective. For low NA-objectives, equation (\ref{eq:h})  simplifies to an expression of the focal shift in the paraxial limit
\begin{equation}
	k(n) = \frac{n}{n_{oil}}.
	\label{eq:k}
\end{equation}
We also compared our measurement to two theoretical studies that take the full vectorial properties of light into account \citep{Hell1993,Sheppard1997}. A summary of these scaling factors is shown in Fig.~\ref{fig:scaling}. The (black) circles are our measurement points, which are connected with a linear fit (dashed black line). The (green) continuous and (green) dashed-dotted lines are from the theoretical prediction of equation (\ref{eq:h}), for NA = 0.7 and NA = 1.4 respectively. The (pink) square is based on a theoretical study by Sheppard \textit{et al.} \citep{Sheppard1997} for NA = 1.4 and the (blue) diamonds show calculations based on a study by Hell \textit{et al.} for NA = 1.3 \citep{Hell1993}, both at a wavelength around 500 nm. The reason for choosing a lower NA in the latter study is that due to total internal reflection at the glass/water interface, a numerical aperture of 1.4 becomes effectively 1.3 \citep{Hell1993}.\\ 
 The calculations by Hell \textit{et al.} seem to agree best with our measurements (black circles). It is also clear from Fig.~\ref{fig:scaling} that the formula based on geometrical optics (equation \ref{eq:h}) is highly dependent on NA and that our measurements do not correspond at all with the theoretical predictions for NA = 1.4. This is a confirmation that indeed the paraxial rays dominate the mechanism of axial shift instead of the high-angle rays used in geometrical optics. Interestingly though, if we assume an `effective NA' of 0.7 (continuous green line), equation (\ref{eq:h}) fits our data remarkably well.

We also measured the axial shift when the calibration cell was filled with CHC and imaged with a 100x oil-immersion objective with variable NA between 0.7 and 1.4 (not shown here). This resulted in an increase in axial distance of 2\% from NA = 0.7 to NA = 1.4, whereas equation (\ref{eq:h}) predicts an increase of 31\%. This result is however again in good agreement with the theoretical prediction and experimental measurement reported by Hell \textit{et al}. \citep{Hell1993}.

From a linear fit to our measurement points, we obtained the empirical formula
\begin{equation}
f(n_D) = 0.82 \; n_D - 0.24,
\label{eq:f}
\end{equation}
with the coefficient of correlation $R^2 = 0.993$ indicating a strong linear correlation. This empirical formula could be used to predict (or estimate) the axial scaling factor for 3D images acquired with an oil-immersion objective (NA = 1.4) for any RI between 1.3-1.5.

\subsection{Calibration with a 50 $\mu$m PMMA sphere}
\begin{figure}[h!]
\includegraphics[width=0.45\textwidth]{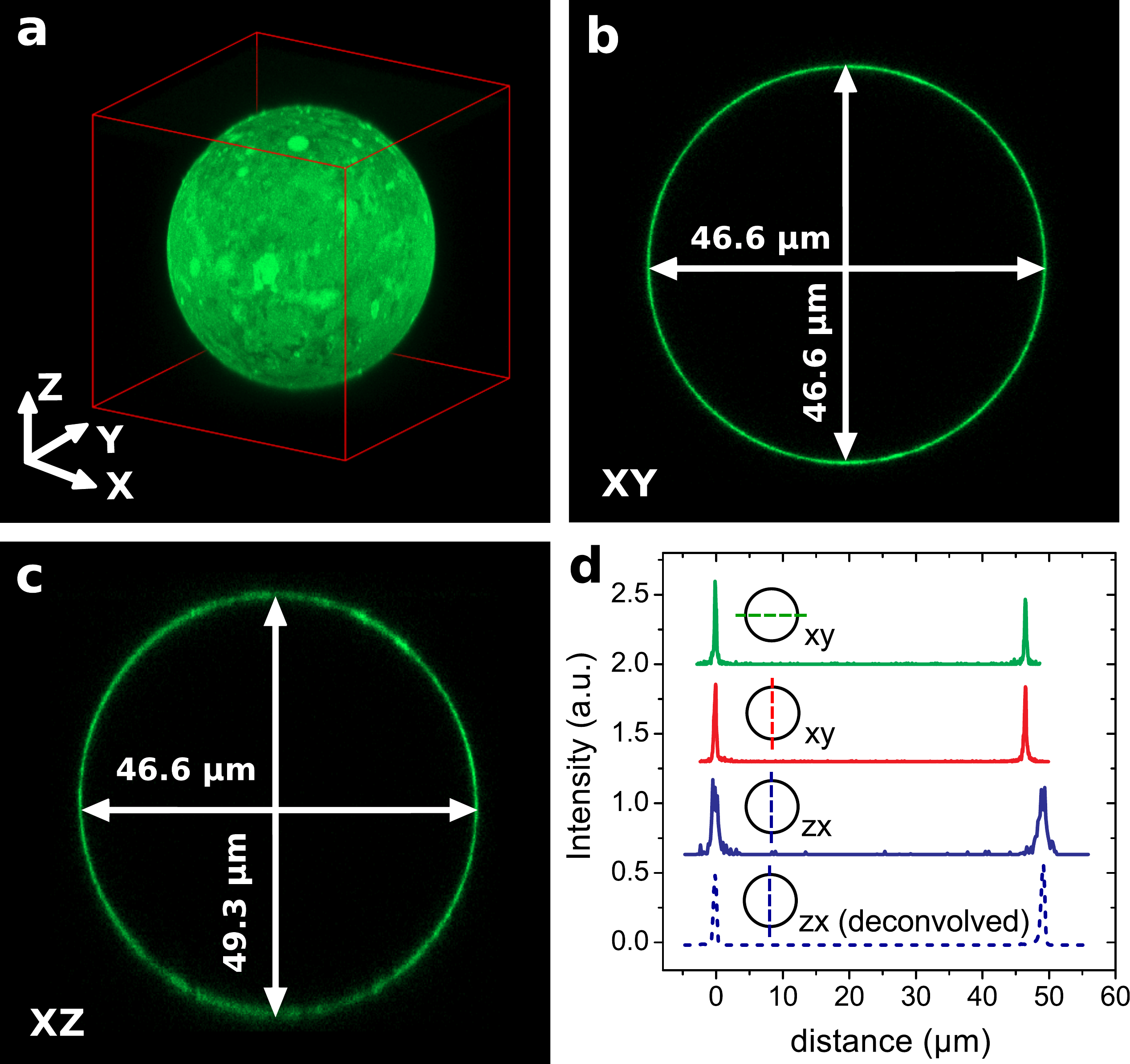}
\caption{A fluorescent PMMA sphere dispersed in an index matching mixture of 24 wt\% cis-decalin in CHB, recorded with a confocal microscope (Leica SP8). (a) 3D view constructed from a XYZ image stack. (b) A single XY image shows that $x$ and $y$ distances are equal. (c) The reconstructed XZ view of the image shows that there is a small (6\%) elongation in the $z$-direction. Due to the refractive index mismatch between the suspension ($n_D$ = 1.49) and the oil immersion ($n_D$ = 1.52) an elongation in the z-direction of 2\% was expected. (d) Intensity profiles along different lines trough the sphere, as indicated in the figure. The profiles were normalized and shifted for better visualization.
\label{fig:PMMA}}
\end{figure}

As a second method to calibrate the axial distance in a confocal microscope, we exploited the well-defined 3D geometry of large spherical PMMA particles (diameter $\sigma$ = 50 $\mu$m and polydispersity $>$ 10 \%), dyed with a thin fluorescent shell ($\sim$ 500 nm). We used these particles to determine the z-calibration of a point-scanning confocal microscope (Leica SP8). We first calibrated the \textit{xy}-distances of the microscope by imaging a calibration grid (Ted Pella, grid spacing 0.01 mm) in reflection mode using a 20x/0.7 air objective (Leica). Then we imaged a single particle in 3D. Fig.~\ref{fig:PMMA}a shows a 3D image-stack of a particle  dispersed in an RI-matching mixture of 24 wt\% cis-decalin/CHB. In Fig.~\ref{fig:PMMA}b, a single \textit{xy}-image shows that the diameter of the particle in the \textit{x}- and \textit{y}-direction is equal. However, a reconstructed \textit{xz}-view of the particle (Fig.~\ref{fig:PMMA}c) shows that there is an elongation in the \textit{z}-direction. From the intensity profiles, shown in Fig.~\ref{fig:PMMA}d, we determined the diameter of the particle in the \textit{x}-, \textit{y}- and \textit{z}-direction, and found an elongation of 5.8\% in the \textit{z}-direction. We also deconvolved the 3D image stack with a theoretical depth-dependent PSF. The resulting intensity profile in the \textit{z}-direction is indicated with the (blue) dashed line in Fig.~\ref{fig:PMMA}d. The deconvolution resulted in a decrease of the width of both peaks, however, there was no significant change in the distance between them. Additionally, we acquired images for different scan-speeds and different image-sizes and found similar results.  
Due to the (small) refractive index mismatch between the suspension ($n_D^{21}$ = 1.490) and the immersion oil ($n_D^{20}$ = 1.516) we expected, based on equation (\ref{eq:f}), an axial scaling factor in the $z$-direction of only $f(1.49$) = 0.98. We therefore conclude that there is a small but significant elongation in the $z$-direction of 3.7\%, which is most likely a calibration error. To confirm this statement, we measured the height of our calibration cell when it was filled with immersion-oil (Fig.~\ref{fig:cell_full}b) with the same microscope and objective as used for the image-stack in Fig.~\ref{fig:PMMA}, and found a distance of $H = 83.4$ $\mu$m. This indicated a similar deviation of 3.0\% in the axial direction. \\ \indent Because the calibration of the \textit{xy}-distances in confocal microscopy is simple and straightforward (e.g.\ with a calibration grid), the fluorescent PMMA spherical particles described above can be used to measure absolute axial-distance deviations within $\sim$~1-2\% and are therefore suitable calibration particles. An additional benefit is that these particles hardly display thermal motion, even when dispersed in a solvent with viscosity $\sim 1$ cP, which is due to their large size.

\subsection{PSF measurement \& imaging of single fluorescent beads}
\begin{figure*}[ht]
\includegraphics[width=0.80\textwidth]{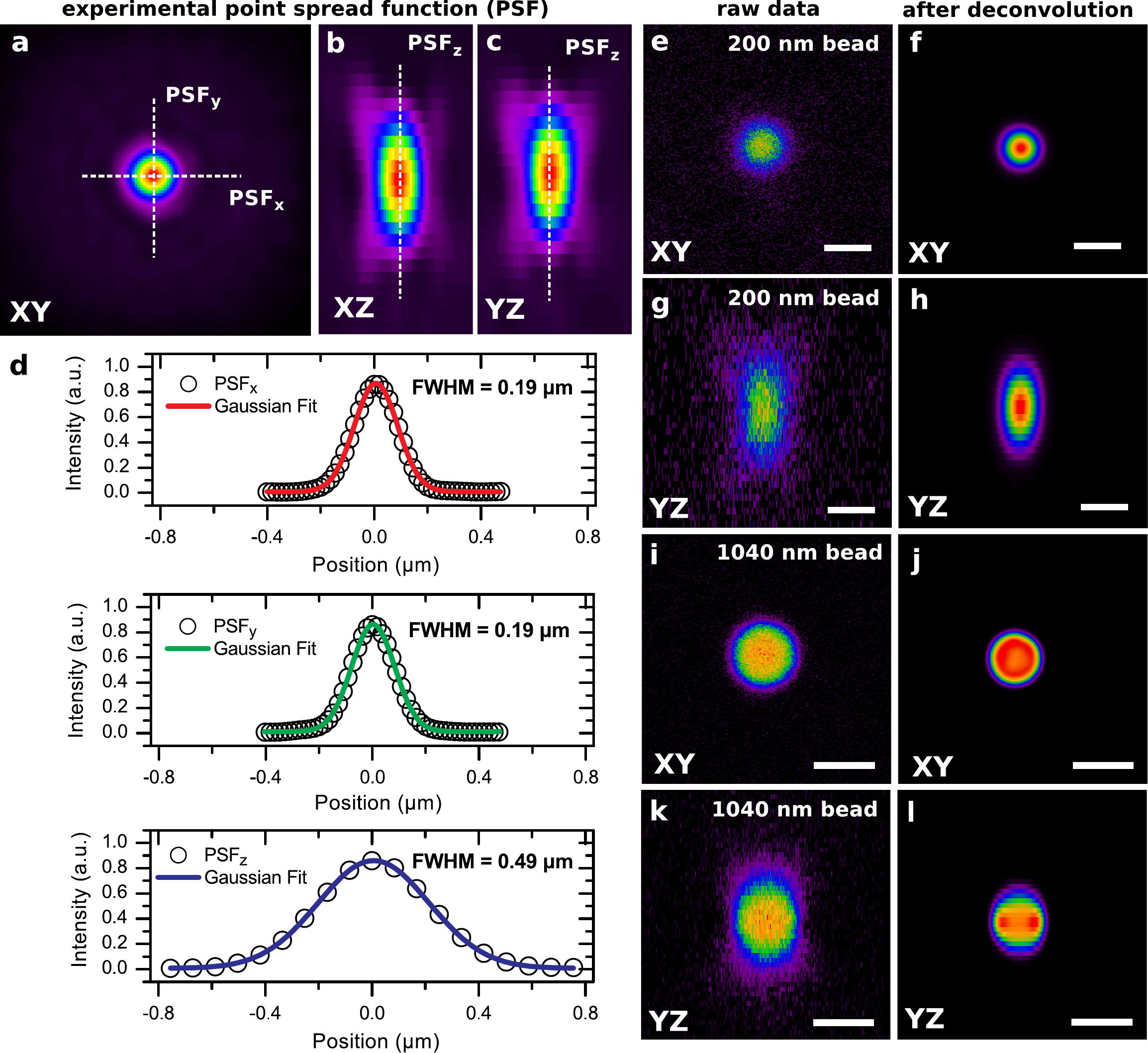}
\caption{Experimental measurement of the point spread function (PSF) and imaging of single fluorescent beads. Images were recorded with a 100x/1.4 oil immersion objective. (a) The PSF in the XY plane. Intensity profiles were recorded along the indicated cross-sections. (b-c) The PSF in the z-direction clearly shows the expected elongation, due to the more limited resolution in the axial direction. (d) Recorded intensity profiles from the images in (a) and (b). The FWHMs that we obtained were 190 nm in the lateral and 490 nm in the axial direction. (e-h) Orthogonal views  of a polystyrene bead with a diameter of 200 nm, before and after deconvolution. The scale bar is 300 nm. (i-l) Orthogonal views of a PMMA sphere with diameter 1040 nm, again before and after deconvolution. The scale bar is 1 $\mu$m. 
\label{fig:PSF}}
\end{figure*}
In Fig.~\ref{fig:PSF} we show examples of an experimental measurement of the PSF and its effect on confocal microscopy measurements of fluorescent particles. In Figs.~\ref{fig:PSF}a-c we show images of the PSF of an  accurately calibrated point-scanning confocal microscope (Leica SP8) equipped with a 100x/1.4 oil-immersion objective (Leica). The intensity profiles of the PSF in the \textit{x}, \textit{y} and \textit{z}-direction could be well fitted with Gaussian functions (Fig.~\ref{fig:PSF}d). From the FWHM of these Gaussian fits, we obtained a measure of the resolution of the microscope. The values that we obtained are 190 nm in the lateral and 490 nm in the axial direction, which is close to the maximum resolution possible for a conventional point-scanning confocal microscope, which is around 178 nm in the lateral and 459 nm in the axial direction for this setup \citep{Cole2011,Wilhelm1997}. Also, the symmetry of the PSF in all three directions is high, indicating little optical aberration. In Figs.~\ref{fig:PSF}e-l we demonstrate the effect of the PSF on the geometry of two (nearly) index-matched spherical particles. In Figs.~\ref{fig:PSF}e-h, orthogonal views are shown of a polystyrene bead with a diameter of 200 nm that was immersed in immersion oil (Type F, Leica) before and after deconvolution. It is clear from Fig.~\ref{fig:PSF}g that its dimensions in the axial direction were stretched. Deconvolution (Figs.~\ref{fig:PSF}f,h) reduced the apparent size of the particle, however, anisotropy in the particle shape still remained. In Figs.~\ref{fig:PSF}i-l, orthogonal views are shown of a PMMA sphere (diameter 1040 nm), before and after deconvolution. Despite its larger size, the particle still seems elongated in the axial direction (Fig.~\ref{fig:PSF}k), however, deconvolution almost recovered the spherical shape of the particle (Fig.~\ref{fig:PSF}l). \\ \indent These measurements of spherical particles demonstrate that even a micron-sized object that was (nearly) RI-matched, seemed elongated in the axial direction due to the anisotropy of the PSF (and possibly to a far lesser extent due to a subtle difference in RI between particle and solvent). This demonstrates that single, micron-sized features are not suitable to determine if the microscope is correctly calibrated in the axial direction, even when the sample is RI-matched.

\section{Discussion}
With the calibration cell described in this paper, we measured the scaling of axial distances as a function of refractive index (RI) mismatch. We found for an aqueous sample dyed with FITC (excitation wavelength 488 nm) imaged with an oil-immersion objective with NA = 1.4, an axial scaling factor of 0.85. This value is in good agreement with the theoretical calculations of Hell \textit{et al.}~\citep{Hell1993}, who found an value of 0.83 and to reasonable extent to the value of 0.89 calculated by Sheppard \textit{et al.}~\citep{Sheppard1997}. The linear slope fitted to our data was however much smaller than the slope predicted from the high-angle geometrical optics equation (\ref{eq:h}), which predicts a scaling factor of 0.36 for NA = 1.3, and is slightly higher than the slope for the paraxial limit $n/n_{oil}$. Our experimental values are however  in good agreement with other experimental measurements that use a fluorescent `sea' between two  coverslips \citep{Hell1993,White1996}. Theoretical expressions that take the vectorial properties of light into account found almost linear scaling in axial shift as a function of axial distance, and also found no strong dependence on excitation wavelength (around 500 nm) \citep{Hell1993,Jacobsen1995,Sheppard1997}, which extends the applicability of these results.
\\ \indent Our measurements deviate considerably however from experimental studies on micron-sized particles that are immersed in a solvent with a RI-mismatch, where scaling factors of 0.4-0.7 are reported for aqueous samples \citep{Visser1992,Visser1994,White1996}. In the case of a RI-mismatch between the sample and the immersion liquid, both the width of the PSF increases \citep{Hell1993,Shaevitz2007}, as well as the apparent axial distance (due to the focal shift). These two effects are hard to separate for micron-sized particles and has led to overestimation of axial distance scaling in previous studies, as described further in Ref.~\citep{Wiersma1997}. The overestimated axial scaling obtained by measuring particles of a few micron in diameter corresponds however approximately to the incorrect axial scaling distances predicted by the geometrical optics model (equation \ref{eq:h}). 
\\ \indent This does not mean that micron-sized spheres are not useful for calibration samples. On the contrary, regular 3D colloidal crystals of fluorescent micro-spheres can act as an ideal calibration sample, because of the well defined (periodic) 3D distances of the crystal lattice. The particles can be immobilized by post-treatment of the sample and the lattice distances can be measured with complementary methods such as light scattering or X-ray diffraction \citep{Thijssen2006}. Such 3D colloidal crystals are especially worth exploring because a complete theory exists on how to correct for refraction index differences between the micro-spheres and the surrounding medium. Presently we are using such samples to test effective medium theories that are used to arrive at approximate effective refractive indices for the combined particle-solvent system. Furthermore, if the particles have e.g.\ a small gold core, the sample can at the same time be used to measure the PSF (in reflection mode).

\section{Conclusion}
We demonstrated two methods to calibrate axial distances in confocal microscopy that are both accurate and practical to employ. The first method consists of a sample cell built from ordinary glass cover-slips. From the Fabry-Perrot fringes in the transmission spectrum of the empty cell, we could accurately measure its height.  We filled the cell with four different solvents mixed with fluorescent dye, which enabled the determination of the axial scaling factors as a function of refractive index for high-aperture confocal-microscopy imaging.  We found that our scaling factors are almost completely linearly dependent on the refractive index (RI) and therefore we determined an empirical formula that provides the axial scaling factor for confocal microscopy images acquired with an oil-immersion objective (NA = 1.4) for any RI between 1.3-1.5. Our results are in good agreement with theories that take the full vectorial properties of light into account, and consequently, there was a strong deviation with the high-angle theoretical prediction of geometrical optics, which predicts much lower scaling factors. The prediction in the paraxial limit (considered only valid for low NA) resulted in only slightly higher scaling factors compared to our measurements, which is in agreement with the assertion that paraxial rays dominate in the mechanism of axial shift. Using a straightforward calibration of the lateral distances of a confocal microscope with a calibration grid, we showed that large ($\sim$~50 $\mu$m) spherical particles that only have a fluorescent shell, can conveniently be used to measure axial-distance deviations and are therefore suitable calibration particles. As an illustration, we demonstrated with a correctly calibrated confocal microscope that spherical objects of only a micrometer or smaller are still significantly elongated due to the PSF, even in the absence of a RI-mismatch, and are therefore not suitable to determine correct axial calibration nor axial scaling.

\subsection{Acknowledgements}
We would like to thank Hans van der Voort, Hans Gerritsen  and Job Thijssen for useful discussion. We thank Johan Stiefelhagen for particle synthesis. This research was carried out partially (THB) under project number M62.7.08SDMP25 in the framework of the Industrial Partnership Program on Size Dependent Material Properties of the Materials innovation institute (M2i) and the Foundation of Fundamental Research on Matter (FOM), which is part of the Netherlands Organisation for Scientific Research (NWO). Part of the research leading to these results has received funding from the European Research Council under the European Union’s Seventh Framework Programme (FP/2007-2013)/ ERC Grant Agreement no. [291667].

\bibliography{./library_confocal_calibration.bib}
\end{document}